\documentclass[useAMS,usenatbib]{mn2e}
\usepackage{graphicx}
\usepackage{amssymb}
\DeclareGraphicsExtensions{.pdf, .jpg}
\usepackage{subfigure,lscape}
\newcommand{\integral}{{\it INTEGRAL}}
\newcommand{\rxte}{{\it RXTE}}
\newcommand{\swift}{{\it Swift}}
\newcommand{\chandra}{{\it Chandra}}
\newcommand{\xmm}{{\it XMM-Newton}}
\newcommand{\beppo}{{\it BeppoSAX}}
\newcommand{\mysou}{{SAX J1753.5$-$2349}}
\def \xmmu {XMMU~J174716.1--281048}
\def \ax {AX~J1754.2--2754} 
\def\ergcms{erg cm$^{-2}$ s$^{-1}$ }
\def\ergs{erg s$^{-1}$}

\def \mdot{$\dot{M}$}
\def \msun{M$_\odot$}
\def \minsim{$\lesssim$}
\def\ks{{\rm{ks}}}
\newcommand{\g}{$\Gamma$}
\title[Hard X-ray spectrum of \mysou]{Unveiling the hard X-ray spectrum from 
the ``burst-only'' source \mysou\ in outburst\thanks{Based on observations with {\it INTEGRAL}, 
an ESA  project with instruments and science data centre funded by ESA member states 
(especially the PI countries: Denmark, France, Germany, Italy,  Switzerland, Spain), 
Czech Republic and Poland, and with participation  of Russia and the USA.}}
\author[M. Del Santo et al.]{M. Del Santo$^{1}$, L. Sidoli$^{2}$, P. Romano$^{3}$, A. Bazzano$^{1}$, R. Wijnands$^{4}$, N. Degenaar$^{4}$,
\newauthor S. Mereghetti$^{2}$\\
$^{1}$INAF/Istituto di Astrofisica Spaziale e Fisica Cosmica di Roma, via Fosso del Cavaliere 100, 00133 Roma, Italy \\
$^{2}$INAF/Istituto di Astrofisica Spaziale e Fisica Cosmica di Milano, via E. Bassini 15, 20133 Milano, Italy\\
$^{3}$INAF/Istituto di Astrofisica Spaziale e Fisica Cosmica di Palermo, via U. La Malfa 153, 90146 Palermo, Italy\\
$^{4}$Astronomical Institute, 'Anton Pannekoek', University of Amsterdam, Science Park 904, 1098XH Amsterdam, The Netherlands}

\begin{document}

\date{Accepted 2010 January 27. Received 2010 January 26; in original form 2010 January 18.}

\pagerange{\pageref{firstpage}--\pageref{lastpage}} \pubyear{2010}

\maketitle

\label{firstpage}

\begin{abstract}
Discovered in 1996 by \beppo\ during a single type-I burst event,
\mysou\ was classified as ``burst-only'' source.
Its persistent emission, either in outburst or in quiescence, had never been observed before October 2008,
when \mysou\ was observed for the first time in outburst.
Based on \integral\ observations,we present here the first
high-energy emission study (above 10 keV) of a so-called ``burst-only''.
During the outburst the \mysou\ flux decreased from 10 to 4 mCrab in 18--40 keV, 
while it was found being in a constant low/hard spectral state.
The broad-band (0.3--100 keV) averaged spectrum obtained by combining \integral/IBIS and \swift/XRT data has been fitted
with a thermal Comptonisation model and an electron temperature $\gtrsim$24 keV inferred. 
However, the observed high column density does not allow the detection of the emission from the neutron star surface.
Based on the whole set of observations of \mysou,
we are able to provide a rough estimate of the duty cycle of the system and the time-averaged mass-accretion rate.    
We conclude that the low to very low luminosity of \mysou\ during outburst
may make it a good candidate to harbor a very compact binary system. 

\end{abstract}


\section{Introduction}

\mysou\ is a neutron star Low Mass X-ray Binary (LMXB) discovered in 1996 by \beppo/Wide Field Camera (WFC)
during a single type-I X-ray burst \cite{zand99}.
However, no steady emission was detected from the source leading 
to an upper limit of about 5 mCrab (2--8 keV) for a total exposure of 300 ks \cite{zand99}.
Cornelisse et al. (2004)  proposed \mysou\ being member of a possible non-homogeneous
class of LMXBs, the so-called ``burst-only'' sources (see also Cocchi et al. 2001).
These are a group of nine bursters discovered by \beppo/WFC
when exhibiting a type-I burst without any detectable persistent X-ray emission.

Recently, \integral\ identified two new members of this class.
In fact, photospheric radius expansion (PRE) bursts have been caught in 
two previously unclassified sources, namely \xmmu\ \cite{brandt06} 
and \ax\ \cite{chelo07}. 
Afterwards, both have been classified as ``quasi-persistent'' Very Faint X-ray Transients (VFXTs), since they
undergo prolonged accretion episodes of many years at low \mdot\ (Del Santo et al. 2007, Bassa et al. 2008). 

VFXTs are transients showing outbursts with low peak luminosity 
($10^{34}$--10$^{36}$ \ergs\ in 2--10 keV),
mainly discovered with high sensitivity instruments on-board
\chandra\ and \xmm\ during surveys of the
Galactic Center region \cite{wij06}.
They are believed to be the faintest
known accretors, and are very likely a non homogeneous class of sources.
A significant fraction ($\sim 1/3$) of VFXTs are X-ray bursters 
(Degenaar \& Wijnands 2009, Del Santo et al. 2007, Del Santo et al. 2008, Cornelisse et al. 2004);
thus they can be identified with neutron stars accreting matter from a low mass companion (M $\lesssim$ 1\msun).

\begin{table}
  \begin{center}
    \caption{Log of the \integral\ observations of the \mysou\ region: 
             orbit number (Rev.), start and end time of the observations, exposures time for each orbit
              taking into account the whole data-set,
             and number of pointings (SCW) are reported. Observations within a single orbit are not continuous.
             The first \integral\ detection of \mysou\ occurred in rev. 732. 
             A data sub-set from rev. 732 to 736 has been used to compute the averaged spectra.
             The last column reports the exposures of spectra in each orbit.}
\scriptsize
    \begin{tabular}{lccccc}
     \hline
      Rev. & Start &  End & Total Exp. & SCW & Spec. Exp.\\
           &    (MJD) & (MJD) &  (ks)     &     &    (ks)\\
      \hline
      724   & 54727.50  & 54728.23 &  58 & 17 & - \\
      725   & 54729.11 & 54731.52 & 198 & 56 & - \\
      726   & 54732.52 & 54734.46 & 160 & 45 & - \\
      729   & 54741.37 & 54741.86 & 42  & 12 & - \\
      731   & 54749.22 & 54749.55 & 20  & 8 & -\\
      732   & 54749.90 & 54750.85 & 83  & 32 & 26.2  \\
      733   & 54754.96 & 54755.46 & 38  & 11 & 10.8\\
      734   & 54756.87 & 54758.54 & 128  & 48 & 36.5\\
      735   & 54760.91 & 54761.53 & 43 & 13 &  23.2  \\
      736   & 54762.03 & 54763.63 & 38  & 49 & 30.0\\
      \hline \\
      \end{tabular}\\
\label{tab:log} 
\end{center}
 \end{table}

In 2002 observations with \chandra\ and \xmm\ allowed to reveal the nature of four \beppo\ ``burst-only'' sources:
one persistent very-faint source, two faint transient systems (with 2--10 keV peak luminosity in the range $10^{36}$--10$^{37}$ \ergs),
and one VFXT (see Wijnands et al. 2006 and reference therein).
For the other five bursters, including \mysou, only the quiescent emission
could be derived ($\sim$10$^{32}$ \ergs; Cornelisse et al. 2004).
Wijnands et al. (2006) proposed these systems, as good candidates to be classified as VFXTs (see also Campana 2009).

\begin{figure}
\centering
\includegraphics[height=6cm]{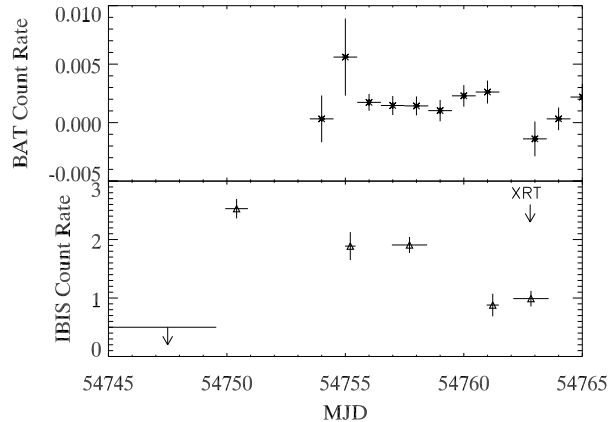}
\caption{\mysou\ BAT (top) and IBIS/ISGRI (bottom) count rate evolution in the 15--50 keV and 18-40 keV energy ranges, respectively.
The XRT detection time is also shown on the bottom plot. The public BAT light curve starts from 54754 MJD; after MJD=54764 \mysou\ was no longer 
pointed by \integral. 
 \label{fig:lc}}
\end{figure}

In 2008 October 11, \rxte/PCA, \swift/BAT \cite{mark08} 
and \integral/IBIS \cite{cadol08} detected an outburst from \mysou\ at 10 mCrab flux level.
Then, \swift/XRT pointed \mysou\ on October 23 \cite{degewij08}, during the decline phase of the outburst (Fig. \ref{fig:lc}).
An improvement in the source position, 
R.A.(J2000)=$17^{h} 53^{m} 31.90^{s}$, Dec(J2000)=$-23^{\circ} 48' 16.7''$, has been provided \cite{starl08}. 
On 2009 March 13, it was re-pointed by \swift\ and a 3$\sigma$ upper-limit derived.
This translates in a luminosity level \minsim $5 \times 10^{32}$ \ergs\ \cite{delsanto09}.  

In this paper we present the hard X-ray outburst of \mysou\ observed by \integral/IBIS,
as well as the first broad-band spectral analysis of the steady emission of a ``burst-only''.
We estimate the long-term mass-accretion rate and discuss the nature of the transient system. 

\begin{figure*}
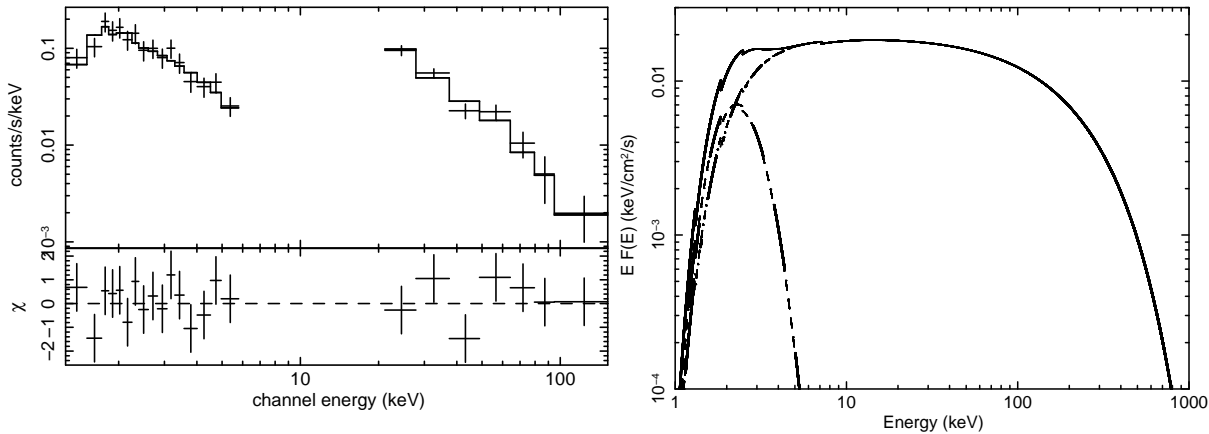

\centering
\includegraphics[height=8cm,angle=-90]{delsanto10_fig2.ps}
\includegraphics[height=8cm,angle=-90]{delsanto10_fig3.ps}
\caption{XRT and IBIS/ISGRI count rate spectra fitted with a simple power-law ({\it{left}}); the total \texttt{bb+comptt} model (continuous line)
 and the two single components (dashed lines) ({\it{right}})
 \label{fig:model}}
\end{figure*}

\section{Observation and data analysis}
\subsection{\integral}
This paper is based on \integral\ observations of the Galactic Centre region carried out in the framework of the 
AO6 Key-Programme. Moreover, we used data from a public ToO on the source H 1743-322,
at 8.6$^\circ$ from \mysou, performed on 2008 October, for a total exposure time of 800 ks (see Tab. \ref{tab:log}).
We reduced the data of the IBIS \cite{ube03} low energy detector ISGRI \cite{lebrun03}, and JEM-X \cite{lund03} data using the \integral\ 
Off-Line Scientific Analysis, release 8.0.
Due to the source weakness, no signal was found in the JEM-X data.
On October 10, the first IBIS detection of \mysou\ was found (rev. 732).
We extracted the IBIS/ISGRI light curves from each revolution as reported in Tab.\ref{tab:log} (binning size as the Total Exposure column)
in the energy range 18--40 keV, 40--80 keV, 80--150 keV.
For the spectral extraction, we used a sub-set of the data reported in Tab. \ref{tab:log}, selecting only pointings
including \mysou\  in the IBIS FOV up to 50\% coding (15$^\circ \times$15$^\circ$).
We obtained four averaged spectra from revolutions 732, 733, 734 and 735-736 (the latests have been added together
because of the poor statistics).
Spectral fits were performed using the spectral X-ray analysis package XSPEC v. 11.3.1.

\subsection{\swift}

A \swift\ ToO was performed on October 23 (Degenaar \& Wijnands 2008).
The {\it Swift}/XRT data of observation 00035713002 
were collected in photon counting (PC) mode between 
2008-10-23 17:48:53 and 21:08:57 UT, for a total on-source net exposure of 1 \ks.

They were processed with standard procedures ({\tt xrtpipeline} v0.12.1), 
filtering and screening criteria by using the {\tt Heasoft} package (v.6.6.1). 
Moderate pile-up was present, so source events were extracted from an 
annular region (radii of 20 and 3 pixels; 1 pixel $\sim 2\farcs36$),
while background events were extracted from an annular region (radii 120 and 80 pixels)  
away from background sources. 
An XRT spectrum was extracted and ancillary response files were generated with {\tt xrtmkarf},
to account for different extraction regions, vignetting and PSF corrections. We used the
spectral redistribution matrices v011 in the Calibration Database maintained by HEASARC.
All spectra were rebinned with a minimum of 20 counts per energy bin. 

We retrieved the BAT daily light curves (15--50 keV)
available starting from MJD=54754, from the \swift/BAT transient monitor (Krimm et al. 2006, 2008; 
http://heasarc.gsfc.nasa.gov/docs/swift/results/transients/) page.

\section{Results}
The IBIS/ISGRI and BAT count rate of \mysou\ are shown in Fig. \ref{fig:lc}.
Based on the IBIS data, the hard X-ray outburst started on October 10
at a flux level of 10 mCrab (18--40 keV) and lasted at least 14 days (last pointing at 4 mCrab). 
This outburst is hence characterised by a fast increase of the flux
and a linear decay with a slope of $-$0.13$\pm$0.01.

An \integral\ pointing with no \mysou\ detection was performed eight hours before 
the outburst started.
We also averaged all our data (from rev 724 to 731) 
collected before the first source detection for a total of 500 ks,
resulting in a 3$\sigma$ upper limit of 1 mCrab (Fig. \ref{fig:lc}).  

In order to look for any possible spectral variability,
we fitted the four averaged IBIS spectra with a simple power law.
We obtained a constant value (within the errors) 
of the photon index (\g $\sim$ 2) which indicates, in spite of the flux variation, a steady spectral state.

The lack of spectral parameter variation led us to average the IBIS spectra of different revolutions.
The 18-100 keV averaged spectrum is well described by a simple power law model with
a slope as $2.2 \pm 0.3$. A mean 18--100 keV flux of $1.5\times 10^{-10}$ \ergcms can be derived.

The XRT spectrum can be fitted by an absorbed power law model
with a Hydrogen column density of N${\rm _H}=1.8 (\pm 0.6) \times 10^{22}$ cm$^{-2}$.
The photon index is $\Gamma = 2.0 \pm 0.5$ and the resulting 2--10 keV absorbed and 
unabsorbed fluxes are $\sim$4.4 and $\sim$5.2 $\times 10^{-11}$ \ergcms, respectively.

We note that the derived  N${\rm _H}$ is higher than the absorption column 
of $0.83 \times 10^{22}$ cm$^{-2}$ \cite{corne02} found by interpolating the HI maps of Dickey \& Lockman (1990). 
In fact, the two values are perfectly consistent within the errors,
given the large range of values (about $0.4-1.5 \times 10^{22}$ cm$^{-2}$) 
obtained in the box adopted to calculate the 
Weighted Average N${\rm _H}$ (with the nH Column Density Tool)\footnote{http://heasarc.gsfc.nasa.gov/docs/tools.html}
from the HI maps. 
 
The joint IBIS and XRT spectrum (0.3--100 keV) was then fitted with different models.
First we used an empirical model such as the power law (Fig. \ref{fig:model}, {\it left}), 
then the more physical Comptonisation model.
Indeed, the 1--200 keV spectrum of X-ray bursters in low/hard state is most likely 
produced by the upscattering of soft seed photons by a hot optically thin electron plasma (i.e. Barret et al. 2000
and references therein). 
Moreover, a black-body emission from the neutron star surface is
also expected to be observed in the low/hard states of bursters (i.e. Natalucci et al. 2000 and references therein).
We tried to add a \texttt{BB} component to the two models.
The best fit parameters and mean fluxes are reported in Tab. \ref{tab:fit_sim}.

\begin{figure}
\centering
\includegraphics[height=7cm,angle=-90]{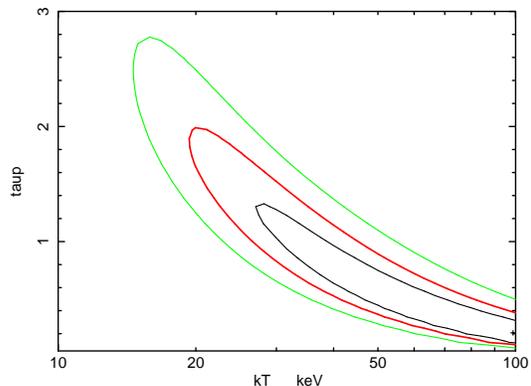}
\caption{Confidence contour levels of electron temperature and plasma optical depth for the {\texttt{comptt}}
model fitting the broad-band spectrum.
 \label{fig:cont}}
\end{figure}

Thus, using a physical thermal Comptonisation model, \texttt{COMPTT} \cite{tita94} in XSPEC,
the electron temperature is not constrained,
while a lower limit of $\sim$24 keV (at $90\%$) can be inferred 
(see Tab. \ref{tab:fit_sim} and contour levels in Fig. \ref{fig:cont}).  
This is consistent with the electrons temperature observed in burster systems,
even brighter than \mysou\ \cite{barret00}.  
 
With the addition of the \texttt{BB} component to the thermal Comptonisation,
a typical value of the black-body temperature (kT$_{\rm BB}$ $\sim$0.3 keV) is obtained (Fig. \ref{fig:model}, {\it right}),
even though this component is not requested by the Ftest probability ($7 \times 10^{-2}$).
We may argue that the high absorption observed in \mysou\ could be a strong obstacle 
to the firm detection of this component.

As a firts approximation, the accretion luminosity L$_{\rm acc}$ is coincident with 
the bolometric luminosity of the source (0.1--100 keV).
Using the mean 0.1--100 keV flux obtained with the \texttt{COMPTT} model fit and
assuming a distance of 8 kpc (Galactic Centre), 
a value of L$_{\rm acc}=4.3\times10^{36}$ \ergs\ ($\sim$0.02 L$_{Edd} $) is derived.
The averaged mass-accretion rate ($\langle \dot{M}_{\mathrm{ob}} \rangle=R L_{\mathrm{acc}}/GM$, 
where $G$ is the gravitational constant, $M=1.4~\mathrm{M_{\odot}}$ and $R=10$~km for a neutron star accretor)
during the outburst is $6.7 \times 10^{-10}$ \msun\  yr$^{-1}$.

\begin{table*}
  \begin{center}
    \caption{The parameters the XRT/IBIS spectra fitted four different models.}
    \vspace{1em}
    \renewcommand{\arraystretch}{1.5}
    \begin{tabular}{lrccccccc}
      \hline
 Model     & N$_{H}$               &  $kT_{BB}$          &  $\Gamma$     & $E_{c}$  &  $kT_{e}$ & $\tau$ & $\chi^2_{\nu}$(dof) & $F_{\rm bol}^{\mathrm a}$ \\
           & $10^{22}$ ($\rm {cm}^{-2}$) &   (keV)             &               &  (keV)   &    (keV)  &            &                  &     (\ergcms)   \\
      \hline
      POW        & 2.2$^{+0.5}_{-0.4}$    &   -                 & $2.3 \pm 0.3$ &   -      &     -     &    -     &  0.91(19)& $1.3\times 10^{-9}$ \\
      BB+POW     & 2.8$^{+2.0}_{-1.0}$    & $0.4^{+0.3}_{-0.1}$ & $2.1 \pm 0.3$ &   -      &           &            &  0.82(17) &  $5.6\times 10^{-10}$ \\
      Comptt    & 1.9$\pm 0.4$          &     -               &   -           &   -      & $> 24$    & $0.2^{+1.3}_{-0.1}$  & 1.07(18) & $1.1\times 10^{-9}$ \\
      BB+Comptt & 2.7$^{+2.0}_{-1.0}$    & $0.4^{+0.3}_{-0.2}$ &   -           &   -      & $> 17$ & $0.8^{+2.2}_{-0.6}$  & 0.86(16) & $6.3\times 10^{-10}$  \\
      \hline \\
      \end{tabular}
    \label{tab:fit_sim}
  \end{center}
\vspace{-0.6cm}
\begin{list}{}{}
\item[$^{\mathrm{a}}$] The bolometric flux of the unabsorbed best-fit model spectrum.
\end{list}
\end{table*}

\section{Discussion}

We report here for the first time the broad-band spectrum,
from soft to hard X-rays, of the persistent emission from
a so-called ``burst-only'' source.
In particular, none of these sources have ever been
studied above 10 keV during their persistent emission.

The outburst from \mysou\ observed with \integral/IBIS has
a duration of at least 14 days, without any evidence for
type-I X-ray bursts, all along the performed \integral\ observations of the Galactic
Centre region started in 2003.

From the \rxte/PCA flux detection at 8 mCrab \cite{mark08}
we can derive an absorbed 2--10 keV peak flux 
of about $1.7 \times 10^{-10}$ \ergcms\ which translates in an unabsorbed 
luminosity higher than $1.3 \times 10^{36}$ \ergs.
This value seems to indicate \mysou\ being a hybrid system (such as AX J1745.6--2901 and GRS 1741.9--2853, 
see Degenaar \& Wijnands 2009) which displays very-faint outbursts
with 2--10 keV peak luminosity $L_{X} < 10^{36}$ \ergs\ (as resulted from WFC observations in 1996),
as well as outbursts with luminosities in the range 
$10^{36-37}$ \ergs, which are classified as faint (FXT; Wijnands et al. 2006).
However, it is worth to know that the $L_{X}$ boundary as $10^{36}$ \ergs\ is
somewhat arbitrary (such as the VFXT/FXT classification).   
Nevertheless, our result reinforces the hypothesis that
the so-called ``burst-only'' sources belong to the
class of the subluminous neutron star X-ray binaries.

A rough estimate of the duty cycle (as the ratio of $t_{\mathrm{ob}}/t_{\mathrm{rec}}$) can be obtained.
The time interval between the two 2008 measurements of the quiescence 
(February 2008-March 2009) is about 13 months
while the outburst recurrence ($t_{\mathrm{rec}}$) is about 12 years (from the burst event in 1996).
However, it is possible that we missed other outbursts of \mysou\ that occurred 
between 1996 and 2008 whithin periods not covered by Galactic Centre monitoring.
The outburst duration ($t_{\mathrm{ob}}$) ranges from a minimum of 14 days (as observed)
and a maximum of 13 months, since there are not any other X-ray 
observations but the ones in October.
In fact, we cannot exclude that the hard X-ray outburst may be part of a longer outburst
occurred at a lower luminosity level, only detectable by high-sensitivity X-ray telescopes.

This translates into a duty cycle ranging from a minimum of 0.3$\%$ to a maximum of 
9$\%$ and into a long-term time-averaged accretion rate ($\langle \dot{M}_{\mathrm{long}} \rangle=\langle 
\dot{M}_{\mathrm{ob}} \rangle \times t_{\mathrm{ob}} / t_{\mathrm{rec}}$)
ranging from 2.2$\times$$10^{-12}$ to 6.0$\times$$10^{-11}$ M$_\odot$ yr$^{-1}$.

King \& Wijnands (2006) suggested that neutron star in transient LMXBs 
with low time-averaged mass accretion rate 
might pose difficulties explaining their existence without invoking exotic
scenarios such as accretion from a planetary donor.
However, the regime of $\langle \dot{M}_{\mathrm{long}} \rangle$ estimated for  
\mysou\ can be well explained within current LMXB evolution models. 

In spite of the flux variability along the outburst,
the spectral state of \mysou\ remains steady, in low/hard state.
This is in agreement with the fact that a really low X-ray luminosity, $L_{X}$ \minsim $0.01 L_{Edd}$ or so, 
produces a hard state in most sources \cite{klis06}.

Following in't Zand et al. (2007), we have estimated the hardness ratio 40--100/20--40 keV
within each \integral\ revolutions. 
We find a value consistent with 1 which confirms the hard nature of the system.
This is also consistent with the low mass accretion rate inferred (see also Paizis et al. 2006), 
i. e. \mysou\ is not a fake faint system and there would be no reason to assume that the system is obscured 
to explain the low \mdot. 

Moreover, King (2000) argued that the faint low-mass X-ray transients are mainly neutron star X-ray binaries
in very compact binaries with orbital periods lower than 80 min.
We suggest that the \mysou\ system is a good candidate to harbor
an accreting neutron star in a very compact system. 

In conclusion, \mysou\ joins a sample of low-luminosity transient LMXBs \cite{degewij09}, which 
display  different behaviour in terms of peak luminosity, outburst duration and recurrence time from year to year.
Up to now, it is not understood whether these variations should be interpreted 
as being due to changes in the mass-transfer rate or as results of instabilities 
in the accretion disc (Degenar \& Wijnands 2009 and reference therein).

\section*{Acknowledgments}
Data analysis is supported by the Italian Space Agency (ASI),
via contracts ASI/INTEGRAL I/008/07/0, ASI I/088/06/0.
MDS thanks Memmo Federici for the \integral\ data archival support at IASF-Roma.
We thank the anonymous referee for his quick response and useful suggestions.

\label{lastpage}

\end{document}